# A Framework for Data Valuation and Monetisation

Eduardo Vyhmeister[1]*, Bastien Pietropaoli[1], UdoBub[2], Rob Schneider[3], and Andrea Visentin[1]

[1]University College Cork, Cork, Ireland
[2]1001 lakes, Helsinki, Finland
[3]FIR e. V. an der RWTH Aachen, Aachen, Germany

*Corresponding author: Eduardo Vyhmeister, E-mail: eduardo.vyhmeister@gmail.com

**As organisations increasingly recognise data as a strategic resource, they face the challenge of translating informational/data assets into measurable and actionable business value. Existing approaches to data valuation remain fragmented, often separating economic, governance, and strategic perspectives and lacking operational mechanisms that can be deployed in real organisational settings. This paper introduces a unified valuation framework that integrates these perspectives into a coherent decision-support model. Building upon two pre-existing artefacts from the Horizon Europe DATAMITE project; the taxonomy of over one hundred data-quality and performance metrics and an Analytic Network Process (ANP) tool for deriving relative importance; we developed a hybrid valuation model. The model combines qualitative scoring, cost- and utility-based estimation, relevance/quality indexing, and multi-criteria weighting to define a data value in a transparent, systematic manner. Anchored in the Balanced Scorecard (BSC), the framework aligns indicators and valuation outcomes with organisational strategies, enabling firms to assess the monetisation potential across diverse pathways such as Data-as-a-Service, Information-as-a-Service, and Answers-as-a-Service. Methodologically, the study follows a hybrid approach of Design Science complemented by embedded case studies conducted with industrial partners in the European Project. The use cases facilitated the continual improvement on the model facilitating users to shape valuation outcomes. Furthermore, since the evaluation is connected to a high-level taxonomy, the approach facilitates information regarding different considerations of the BSC. Based on use cases evaluation and constructed exemplifications, the framework showed to possess flexibility and transparency, reduces arbitrariness in valuation decisions, providing a structured basis for organisations to connect data assets with strategic and economic outcomes.**

# 1 Introduction

Organisations increasingly view data as a strategic asset that fosters competitive differentiation and drives innovation in today's digital landscape. Across industries, companies are investing heavily in collecting, storing, and processing information, elevating it to the same level as traditional assets [1]. Unlike physical property, whose value often diminishes over time, data can gain value through repeated use, provided that it meets quality and timeliness criteria [2]. As data grows in strategic importance, a persistent challenge emerges: how to define and quantify its actual value.

There is no universal methodology for valuing data or defining data monetisation [1, 2]. While some organisations focus on direct sales, others use data internally to enhance efficiency or innovation, employing approaches that range from assigning financial prices to assessing strategic impact. In the absence of formal guidelines, firms adopt varied, often inconsistent methods, with value drivers such as utility, quality, and scarcity remaining difficult to operationalise within a cohesive framework. As a result, understanding data value requires looking beyond isolated valuation techniques and towards the broader organisational and ecosystem conditions that shape monetisation outcomes.

Foundational work in governance and asset-orientation further clarifies the conditions under which data monetisation models succeed. [3] argues that robust data governance, covering decision rights, roles, and accountability, forms a necessary precursor for treating data as an asset capable of being monetised. Similarly, [4] explores how data-portability mechanisms within platform ecosystems influence the competitive dynamics of data-monetisation business models. Additionally, [5] demonstrate how token-based value exchange mechanisms can reshape how data assets are packaged, traded, and monetised. Collectively, these insights from the Information System field highlight that data monetisation is not purely a matter of valuation formulae, but also requires appropriate governance, architecture, and business-model innovation.

Recent contributions have advanced the discussion by introducing algorithmic and explainable mechanisms for quantifying data value in marketplace settings. For instance, [6] propose a Data Asset Value Quantification and Selection Mechanism (DQSM), which leverages Machine Learning and Explainable AI to evaluate the potential utility of data assets without requiring providers to disclose raw data. Their framework could help mitigate the asymmetry between data buyers and sellers, thereby enhancing efficiency and trust in data marketplaces [6]. Such mechanisms offer a practical complement to strategic valuation frameworks by operationalising data value estimation through automated modelling. However, despite their practical relevance, mechanisms such as the DQSM depend heavily on the availability of labelled data and suitable model references, which may not exist in early stage or exploratory monetisation settings. Second, these approaches require assumptions about context, data completeness, and feature-label relationships; when these conditions do not hold, the resulting value scores may be unstable or misleading.

What is needed is a coherent approach that bridges the gap between assigning value to data and translating that value into tangible business outcomes. For example, [7] identify two categories to determinate data value; Essential Factors (completeness, accuracy, uniqueness, consistency) and Value-of-Use factors (risk, timeliness, accessibility, usage restrictions, utility). This hierarchical delineation could facilitate the quantification and operationalisation of data-valuation approaches.

Recent work [8] has emphasised that before organisations can even define valuation or pricing frameworks, they must understand their own maturity level in data monetisation. In fact [8] introduce a Data Monetisation Maturity Model (DMMM) that helps firms to assess their current capabilities, identify organisational gaps, and prescribe target actions toward more advanced monetisation practices.

From an enterprise perspective, data valuation is shaped not only by hierarchy but also by strategic priorities. Firms focused on differentiation may value data that enhances customer insight and innovation, while those pursuing cost could prioritise efficiency and cost reduction. Aligning valuation frameworks with strategies helps translating business criteria into technical metrics, facilitating the valuation of high-impact datasets and positioning data as a core commercial asset.

In response to these challenges, this study introduces a design-science-inspired framework that consolidates existing elements of data valuation into a coherent structure suitable for organisational use. Instead of proposing an entirely new valuation paradigm, the contribution lies in integrating established components, namely a metrics taxonomy [9] an Analytic Network Process (ANP) for decision support [10, 11], and cost/utility-oriented valuation elements, into a unified and operational model. Furthermore, the framework positions data valuation as a practical decision-support activity that can be aligned with business priorities through the Balanced Scorecard (BSC) [12] perspectives, ensuring that valuation outputs remain transparent, traceable, and strategically grounded.

Rather than claiming completeness or universality, the framework offers a structured way for organisations to navigate large sets of heterogeneous metrics, derive context-specific priorities, and compute interpretable value estimates. Its purpose is to support managerial decision-making by clarifying how different dimensions (e.g. quality, relevance, utility, governance, and cost) interact within real settings.

The framework is supported by a multi-domain case-study methodology that helps to extend its general applicability. These case-studies are contained within the European Project DATAMITE that helped shape the framework applicability under different contextual constraints, leading to a refined model through practitioner feedback.

The manuscript is structured as follows: Section 2 introduces key concepts in data valuation and explores the relationship between strategies and data valuation models. Section 3 present the Research Design and Methodological Positioning, a description of the protocol followed in the evaluation, and the proposed framework. Section 4 describe in detail the framework presented in this work, including the mathematical formulation for estimating the data assets value. Section 5 illustrates the framework's application through examples and defines the domains in which the model was tested. Finally, Section 6 summarises the main findings and outlines directions for future work.

## 2  Background

In recent years, a growing body of literature has sought to establish systematic methodologies for assessing the value of data and for deriving appropriate pricing schemes when data are exchanged in market contexts. These efforts can be broadly classified into two clusters [1]: **Internal Valuation** methods and **External Pricing** methods [13].

Internal valuation approaches assess the intrinsic, operational, and strategic value of data within an organisation, often yielding qualitative scores or estimates of cost savings, risk reduction, or expected gains. These methods typically emphasise governance factors, such as data quality, compliance, and usage, and rely on internal cost figures to guide management decisions. However, they do not produce explicit market prices suitable for external transactions [1].

In contrast, external pricing methodologies are explicitly concerned with setting prices for data products in a marketplace [1]. Some introduce mechanisms to ensure arbitrage-free pricing across queries or subsets and distribute value among individual records or data contributors, often starting from a baseline value, derived through cost-based or utility-based anchors. The estimation can be further adjusted based on scarcity, timeliness, or buyer-specific customisation. Fundamentally, the anchoring can follow a market benchmarking, pricing datasets by comparison with similar offerings.

In practice, the appropriateness of each valuation or pricing method is determined by the firm's strategy: efficiency-focused organisations favour cost- or utility-based approaches, while market-oriented strategies could require methods that capture external demand. Thus, a model is valid only insofar as its assumptions align with the strategic context. Importantly, both approaches are not necessarily mutually exclusive and can be aligned with broader classifications such as cost-, income-, market-, intrinsic-, or outcome-based approaches [2].

## 2.1 Strategies and Monetisation Approaches

Data monetisation strategies fall into two categories: direct, where revenue is generated through explicit sales or licensing of data products, and indirect, where value is captured through enhanced services, improved decision-making, or operational efficiencies [13]. Certain valuation and pricing methods span both categories. For example, internal utility-oriented estimates can inform cost setting and market pricing (direct) as well as guide investment prioritisation for efficiency gains (indirect). Likewise, cost-based measures may underpin subscription fees or support infrastructure investments that enable downstream analytics.

Building on the direct/indirect distinction, organisations can deploy a diverse spectrum of data-monetisation strategies; ranging from traditional models like Data-as-a-Service (DaaS), Information-as-a-Service (IaaS) and Answers-as-a-Service (AaaS), to approaches focused on cost reduction, data-network effects, wrapping, servitisation, launching new products or services, data bartering, and platform-provider or platform-refiner [9, 13, 14].

Recognising the connection between strategies and valuation in a dynamic environment allows organisations to build cohesive strategies. In fact, firms rarely follow a single monetisation pathway; they may sell raw data, optimise operations, or introduce new models as markets and technologies evolve. Therefore, a robust framework for valuation must be modular, enabling businesses to recalibrate their evaluation in response to shifting customer needs, regulatory pressures, and competitive dynamics.

## 2.2 Data Valuation Model Types

As observed in the literature, different frameworks can be used to classify data valuation models. From the perspective of [15], the methods can be classified depending on three dimensions:

- Input Type: metadata-based, data based, and bimodal.

- Use Context: decision-making, and content, offline, and online utility analysis.

- Methods: multicritera decision-making, mathematical modelling, and machine learning.

Following this logic, our work could be classified as a hybrid approach since no combination represents internal valuation metrics, strategic alignment, and external pricing logic within a single coherent framework.

**Internal Valuation Methods:** Internal valuation methods [1, 16, 17] can be succinctly grouped into three overarching categories:

*Qualitative Scoring*: These methods assign datasets qualitative scores reflecting quality, compliance, and strategic importance [16, 17]. Based on [1], approaches for data valuation within this group include manual or automatic assessments [18], assessment survey [16], information management lifecycle [19], and knowledge graphs and maps [20, 21]. Qualitative Scoring methods excel in governance and stakeholder communication contexts, as they translate technical attributes into intuitive metrics.

From the perspective of the taxonomy of Ebiel et al. [15], qualitative scoring corresponds primarily to metadata- or bimodal-driven, content analysis or decision-making valuation methods, often implemented using multi-criteria decision-making (MCDM) tools such as weighted-sum models, fuzzy scoring, or rule-based assessments.

*Cost/Income-Based*: These methods sum up historical and ongoing acquisition costs, cleaning, storage, and maintenance to establish a baseline proxy value. For instance, this group includes Cost-Based Approaches [22] and File Retention Value Calculations [23]. Cost-Based Measures are highly useful for establishing a defensible baseline value rooted in actual expenditures, which aligns well with financial reporting and budgeting processes.

*Utility-Oriented Estimates*: These models project economic benefits, utility quantitative values, or opportunity costs by linking the dataset attributes to different set of contextual variables such as data usage, risk mitigation, operational gains and competitive advantage. Within this group we can have approaches such as metric-based approaches (e.g. factor assets pricing valuation or Data Quality valuation) [7, 24–26], and governance and the opportunity / risks involved within data [17, 27, 28].

This category maps closely to the offline and online utility-analysis contexts in the taxonomy from Ebiele et al. [15], where valuation is determined by the contribution of data to a predefined analytical or operational goal (e.g., accuracy, uncertainty reduction, performance improvement).

Utility-Oriented Estimates often provide the most strategic insight by directly linking data to projected benefits, making them invaluable for return on investment (ROI) or quantitatively-focused decision-makers.

**External Valuation** When referring to external valuation, different methods can be described. For example, query-based pricing interpolates arbitrage-free prices for any requested subset from a few reference prices already set for a whole dataset (but offers no guidance on how to establish those base prices) [29]; tuple-based pricing takes a known overall dataset price and apportions it fairly among individual records using value drivers (e.g. data cost, entropy, relevance, source credit, reference index), yet likewise presumes an existing total price [1]; simulation approaches [14, 30, 31] use market representation or other frameworks to model the strategic interactions between data buyers, sellers and brokers, helping to estimate prices, demands, or analyzing market trends.

Since these approaches rely on a predefined baseline price or demand, whether for a dataset or a set of queries, they function as benchmarking-based methods, deriving prices by extrapolation from reference points.

**Hybrid Frameworks** There also exist hybrid frameworks that yield consolidated scores or data value proxies that inform governance, investment decisions, and risks. For example, [21] combine knowledge graphs with metric-based approaches to reflect market trends (i.e. connecting internal valuation methods with external indexes).

Stander's Decision-Based Valuation (DBV) [32] is a hybrid framework that combines cost/income measures, utility-oriented estimates and qualitative governance into a decision-centric method. At its core, DBV decomposes information into "Decision Nodes," each characterised by eight attributes grouped into four dimensions:

- Structure Dimension: *Decision*, which specifies the managerial choice being supported, and *Information Requirements*, which detail the precise data inputs and formats needed.

- Value: *Value Range*, defining the minimum and maximum benefit that can produce, and *Lifecycle*, indicating the period during which the decision remain valid.

- Quality: *Frequency*, denoting data refreshing and timing variance, and *Accuracy*, specifying the required precision levels.

- Thresholds: *Maximum Cost*, the upper limit for data acquisition, and *Return on Investment*.

Once these attributes are defined, DBV allocates all relevant costs to establish a defensible baseline valuation.

From the standpoint of the [15] taxonomy, hybrid models such as DBV do not align neatly with any single branch. [15] emphasise that existing quantitative methods remain siloed, either metadata-driven, data-driven, or bimodal, and tied to narrow contexts such as decision-making or ML-based utility estimation.

No current quantitative method integrates organisational governance, strategic objectives, financial considerations, and utility-based metrics into a single coherent framework. This underscores a gap for methodologies that can bridge internal valuation logic with strategic and market-oriented valuation signals.

## 3 Methodology

### 3.1 Research Design and Methodlogical Positioning

Recent systematic evidence shows that existing quantitative data-valuation approaches are highly heterogeneous across domains, using incompatible definitions, dimensions, and computational techniques, which "makes a cross-domain understanding challenging" [15]. In that study, it is identified that 48% of all valuation methods originate from machine learning contexts, where value is operationalised as a dataset's marginal contribution to model accuracy, convergence, or robustness.

As observed from their work [15], no existing approach combines internal valuation metrics, relative importance weighting, strategic alignment, and external pricing logic within a single coherent framework. This absence of integrative models reinforces the need for approaches that can bridge governance, operational value, and monetisation potential.

This study adopts a hybrid Design Science Research (DSR) and embedded case-study methodology, consistent with guidance from [33], [34], [35], and [36]. Unlike traditional DSR studies that involve the creation of new artefacts, this work focuses on the instantiation, integration, and evaluation of existing artefacts developed within the Horizon Europe DATAMITE project in order to construct a general artefact (i.e. a framework) for data assets valuation.

In the sense of [33], this work follows the core DSR principles (relevance, rigour, and design as an artefact). The relevance is satisfied because the research is grounded in concrete organisational needs and real information provided by DATAMITE industrial partners, resulting in a framework that addresses existing and well-defined practical problem. The rigour is reflected through the use of a pre-specified protocol for framework improvement and the use of validated artefacts (i.e. taxonomy and a multi-criteria decision tool (ANP)). The framework adheres to [33] definition of an artefact as a purposeful, problem-solving construct. In this case, the integrated data valuation framework that operationalises quality, relevance, utility, cost, and demand into a coherent decision-support model.

Furthermore, the role of this paper is aligned with the contribution types described by [34]. Specifically, the study constitutes a situated instantiation and improvement rather than the creation of an entirely new class of artefact. The taxonomy and ANP tool are existing artefacts whose utility has been broadly demonstrated, inlcuding within DATAMITE. Instead, this work adopts them as foundational components for constructing a unified data-valuation framework.

Thus, the contribution of this paper lies in their composition into a new, more generalised framework and in the evaluation of this integrated artefact in multi-domain organisational settings. By demonstrating how these artefacts interact, and by extending them into a hybrid valuation model that can be applied across multiple monetisation strategies, the paper provides the type of integrative and contextually grounded design knowledge that [34] emphasise as foundational for impactful design research.

The embedded case-study component of our methodology follows the principles articulated by [35] and [36]. In line with [35], the DATAMITE industrial cases represent heterogeneous organisational contexts, monetisation objectives, and data-quality conditions, each comprising multiple units of analysis including organisational setting, dataset characteristics, and metric selection, resulting in different valuation outputs. Each case was examined through the structured within-case protocol, followed by cross-case comparison to assess consistency, contextual sensitivity, and areas for model refinement.

Therefore, from a methodological positioning, our study:

- Integrates pre-existing DATAMITE artefacts (taxonomy [9] and ANP tool [11]) into a new hybrid valuation model combining quality, relevance, utility, cost, and demand.

- Instantiate and operationalise the integrated model in real organisational contexts.

- Performs empirical model evaluation using embedded multi-case studies, involving DATAMITE industrial partners operating in data-intensive domains.

Case studies are appropriate because data valuation is deeply contextual and shaped by organisational processes, governance structures, risk profiles, and domain-specific requirements.

The combination of DSR (artefact creation) and case study (artefact evaluation) ensures methodological rigour and relevance. This hybrid methodology is also aligned with emerging Information Systems research on data governance, valuation, and data-driven business models, where analytical models must be tested within operational settings to ensure validity and usefulness.

### 3.2 Methodological Protocol

The six-step methodological protocol used in this study draws jointly on [35] theory-building logic and [36] systematic case-study procedures. From [35], we adopt theoretical sampling, using heterogeneous DATAMITE industrial partners to ensure purposeful variation across organisational contexts, strategies, and data-quality conditions. Cross-case comparison, were used to identify convergent patterns and boundary conditions of the framework. From [36], we incorporate the principles of embedded case designs, treating each use case as comprising multiple units of analysis (organisational setting, dataset characteristics, metric selection, ANP weighting outputs, and valuation outcomes). Structured steps were defined for applying the framework consistently across cases and used analytic generalisation.

The protocol is as follows :

1. Problem identification and contextual grounding.

2. Artefact scoping and objective definition. Specifies the purpose, boundaries, and intended use of the valuation framework, including its required components and assumptions.

3. Model instantiation. Integrates taxonomy, ANP weighting, and valuation logic in function of use cases information.

4. Demonstration through embedded cases and examples. Applies the instantiated model to multiple real-world use cases and structured examples to illustrate how it operates in practice.

5. Evaluation via within-case analysis, cross-case comparison, and iterative refinement.

6. feedback and framework improvement.

### 3.3 Artefacts Justification and Development Process

The proposed framework builds on three interlinked artefacts:

- A metric and KPI taxonomy, already constructed within the DATAMITE project and which can be found in [9, 37].

- DATAMITE ANP-based decision-making tool [11].

- the hybrid valuation model developed in this paper.

The integration of the DATAMITE taxonomy and the ANP-based decision-support tool is grounded in both theoretical and practical considerations. The taxonomy is employed because data valuation requires a multidimensional and conceptually coherent representation of the factors that influence data value-quality, utility, relevance, governance, operational efficiency, and strategic fit. The taxonomy provides a structured, Balanced-Scorecard–aligned hierarchy that organises more than one hundred metrics into stable clusters and sub-clusters. This structure enables systematic indicator selection, avoids arbitrary metric choice, and ensures that valuation inputs reflect established organisational performance perspectives. The taxonomy therefore functions as the conceptual scaffold for the valuation model, allowing different monetisation strategies to be mapped to relevant KPIs in a transparent and replicable way.

The ANP is employed in this framework because it enables the extraction of relative importance weights among interdependent metrics, which is essential for integrating heterogeneous indicators into a unified data-value estimation. Furthermore, in data valuation, metrics influence one another and cannot be treated as independent, additive components. ANP's capacity to model these interdependencies and derive transparent, defensible priority vectors ensures that each metric contributes proportionally to the final valuation in accordance with strategic objectives and expert knowledge. Within the DATAMITE use cases, ANP was therefore used to translate expert judgements into a structured weighting scheme that directly feeds into the valuation model, enabling organisations to consistently "ponder" or modulate the overall data value based on what matters most for their chosen monetisation strategy.

The rest of this section explains the rationale for using each artefact, their own development, their role within the framework, and how they were integrated and instantiated for evaluation. A full coverage of the framework has been set in a different section, assuring the understanding of the novelty of the work performed in this work.

### 3.3.1 Taxonomy

Effective data monetisation starts with tracking performance using clear metrics and KPIs. These translate abstract goals (like faster decisions, better customer insights, or new revenue) into measurable outcomes, allowing organisations to monitor progress, identify gaps, and adjust investments in real time. Without this quantitative grounding, it is impossible to assess whether data initiatives deliver strategic or financial value across use cases, units, or time frames.

By linking these metrics to specific monetisation strategies via a unified taxonomy, it is possible to unlock a generalisable framework for data valuation. Each data-driven strategy, whether direct monetisation, indirect monetisation, or hybrid models, can be assigned to a set of customised indicators.

This taxonomical approach not only clarifies how different strategic levers drive value but also provides a modular template that organisations can adapt and use as their data maturity and market conditions change.

The taxonomy is used for two reasons. First, data valuation requires a multidimensional view that considers contextual factors; a taxonomy provides the necessary conceptual completeness and consistency. Second, the taxonomy enables comparability across datasets and organisations by providing a stable reference structure against which valuation components can be selected, filtered, and prioritised.

Within the present study, the taxonomy serves as the entry point for identifying candidate indicators based on strategy and context, before they are weighted and operationalised in the valuation model.

Within the DATAMITE project [37] was developed a comprehensive taxonomy that organises hundreds of metrics and KPIs into coherent enterprise perspectives. Specifically, it builds on the well-established Balanced Scorecard framework to group and align these indicators across BSC dimensions. This structure facilitates the progression from raw data and process indicators, through appropriate data valuation techniques, toward actionable data valuation. The resulting framework captures the variety of the different strategies, offering a generalisable, modular template for data valuation.

Figure 1 shows a higher-level view of the taxonomy in which a descriptive linkage of the different clusters used for the taxonomical representation and hierarchical dependencies of metrics and KPIs. For more detail on the taxonomy, the reader is encouraged to review [9, 37].

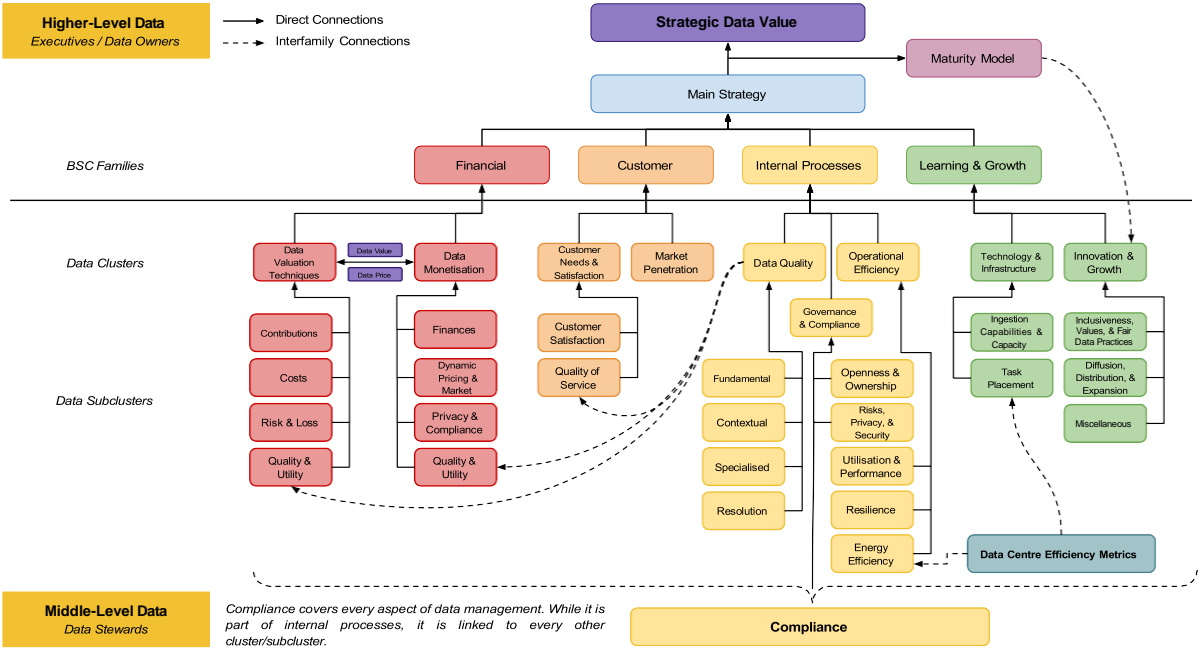

Figure 1: Taxonomy - A Higher Level Visualisation and the Connection to the BCS Components

Within the taxonomy, there are three main metrics / KPIs that are used within the framework to rationalise the value of a dataset. Their nature and connection to an overal context view of the dataset can be understood as follows:

**Data quality:** Data Quality can be seen as the degree to which the data correctly and completely represent the values of the real world and are consistent with the required standards. High-quality data enable reliable insights, support accurate pricing or valuation, and underpin the trust of buyers or internal stakeholders. In practice, data quality is commonly structured into a set of key dimensions. For instance, [38] identify dimensions such as Accuracy, Accessibility, Appropriateness, Believability, Clarity, Completeness, Conciseness, Currency, Equivalence, Redundancy Equivalence, Estimation, Precision, Timeliness, Understandability, and Volatility. Building on this foundation, other authors have developed comprehensive taxonomies and frameworks dedicated specifically to defining metrics for these dimensions of data quality [39, 40].

The considerable number of dimensions / metrics that can be used to track data quality are not necessarily independent. Therefore, an ideal or minimum number of them, as fostered in the framework, could be defined based on the context or strategy for data monetisation.

**Relevance:** Relevance in data valuation denotes how effectively a resource, such as a dataset or system, advances specific organisational goals. By focusing on an organisation (i.e. context), relevance measures how well a resource underpins objectives like decision-making, cost optimisation, or performance enhancement. Importantly, relevance can be quantified as a KPI, enabling direct comparison and trade-off analysis [41].

**Utility:** Relevance measures how well data serves a specific user, use case, or context, whereas Utility captures its broader, intrinsic potential. Thus, data may score high on Utility-being broadly applicable-but still lack Relevance if it doesn't meet a particular need. Conversely, data perfectly tailored to one niche may be highly Relevant yet offer little Utility elsewhere. In this way, Utility can be viewed as the aggregate of Relevance across different domains.

### 3.3.2 ANP-based decision-making tool

The ANP is employed in this work because data valuation inherently involves interdependent criteria, such as quality, relevance, utility, governance, cost, and strategic alignment, that cannot be represented through simple hierarchical models. ANP is well suited to such socio-technical decision problems: it captures mutual influences between indicators, integrates expert judgement into transparent and defensible weights, and incorporates consistency checks that strengthen the reliability of the weighting process. Prior applications of ANP in KPI selection and performance measurement show its effectiveness in contexts where criteria interact and organisational priorities must be reflected in the resulting weights [42]. These characteristics align closely with the requirements of organisational data valuation, where indicators are interconnected and valuation must remain strategically grounded.

Within our framework, the relative importance of each metric is derived through ANP applied in the context of the DATAMITE industrial use cases. By modelling metrics and KPIs as interdependent nodes, ANP accounts for both direct and indirect relationships across governance, cost, utility, and strategic dimensions, ensuring that prioritisation reflects each organisation's monetisation goals. To support this process, the DATAMITE project provides a dedicated ANP-based decision-support tool[1], which guides users through survey-based pairwise comparisons and computes priority vectors that feed directly into the composite valuation model. This ensures that the final data-value estimates are systematically aligned with organisation-specific strategies.

## 4 Framework

The methodology adopts a hybrid valuation framework in which each data-monetisation strategy is tightly interlinked with its requisite components and informed by our unified taxonomy of metrics and KPIs. For any given strategy first, the most relevant indicators are selected from the BSC-based taxonomy (governance, operational efficiency, customer value, innovation, etc.), and then align them with quality, relevance, or utility perspectives. These strategy-specific metrics and KPIs are merged together with cost- or income-based proxies to produce a composite estimate of the value of the data. In practice, this means mapping high-level objectives to measurable inputs (e.g. data-quality scores, usage frequencies, provider credit ratings), applying the appropriate valuation lens (quality, relevance, or utility), and then aggregating the results into a decision-centric value proxy that can directly inform governance, investment prioritisation and risk management.

---

[1]https://datamite.insight-centre.org/

## 4.1 Valuation Framework and Valuation Model

To support a clear understanding of the proposed valuation approach, Figure 2 provides a high-level schematic of how its core components interact. As shown, the approach consists of two main elements: (1) a valuation framework, which defines the process for combining predefined methods with external information to generate an estimated data value, and (2) a valuation model (embedded within the framework), which applies structured indexes from the framework, together with pre-filtered context, to perform the mathematical estimation of the dataset's value.

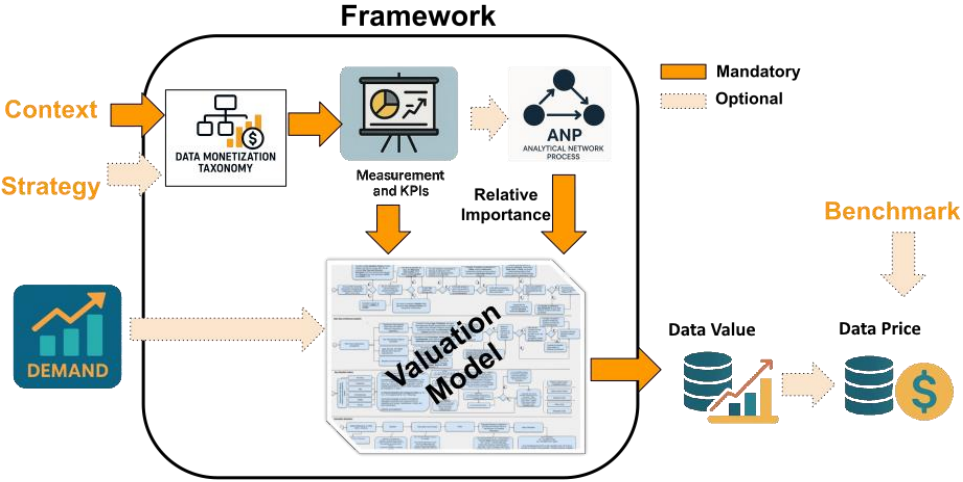

Figure 2: Higher Level Valuation Process Schematic

The valuation framework relies on three main inputs: (1) the enterprise's strategy (e.g., focusing on customer acquisition, operational efficiency, or innovation), (2) the specific operational context of the dataset (including regulatory requirements, competitive landscape, and technological capabilities), and (3) the demand for the service. Of these, strategy and context (where context must be explicitly defined) guide the selection of the most suitable branches within the taxonomy, ensuring that chosen metrics align with the organisation's objectives and environment. Demand estimation, while optional and strategy-dependent, supports the assessment of the dataset's relative value within the framework. Here, demand refers to the goods or services derived from the data, which may differ from the dataset itself.

As seen in the figure, the taxonomy process defines a set of relevant metrics and KPIs to be estimated. To obtain the most suitable metrics and KPIs, the framework can follow two alternative paths. In the first path, users apply decision-making methodologies to determine the relative importance of each indicator. Using multi-criteria decision-making techniques (such as the ANP) a weighted ranking is generated, reflecting expert judgment on the significance of individual data attributes in relation to the organisation's goals. As noted earlier, this entire metrics selection and ANP process has already been operationalised.

The second path, seen as a fallback pathway, involves a decision-tree mechanism, embedded as the initial step in the Valuation Model. Here, the decision tree guides users through a structured series of diagnostic questions, ensuring that at least a minimal set of critical attributes, such as data completeness, timeliness, and volume, are always considered. Unlike the ANP approach, this path does not automatically assign relative importance to indicators. Instead, users must define direct proxies for importance when needed.

This fallback pathway is particularly valuable in contexts where the formal taxonomy cannot be directly applied. For instance, with novel data sources, limited prior experience, or rapidly changing markets. By embedding the decision tree, the framework ensures that minimum quality dimensions and performance indicators are consistently addressed, regardless of the organisation's level of data maturity.

In this framework, the context is generated either through expert knowledge captured via the ANP - taxonomy approach or by extracting it using the decision-tree mechanism. Both processes ensure that the contextual information required for valuation is systematically defined and available for use in the model.

Independently of the path selected, in the valuation model, the metrics and their relative importance are combined with external modifiers to produce valuation estimates. More specifically, the demand and cost estimations serve to amplify the dataset's value determination. The demand estimation approach is out of the scope of the present work. Nevertheless, readers are encouraged to review approaches such as [14] for further information.

With the selected metrics, context information, and estimated indexes in place, the Valuation Model can then be applied. This model combines these inputs (context, indicator values, and relative importance) through an approach inspired by the Data Base Valuation (DBV) method [32]. While our formulation builds on DBV principles, it has been substantially reworked to ensure generalisability across different strategies. The outcome of this process is an estimate of the dataset's relative value (see Valuation Model output in Figure 2).

Although not mandatory, this output can also serve as a proxy for benchmarking. In practice, such benchmarking strengthens the valuation by providing a defensible pricing reference that organisations can use in negotiations, investment decisions, and strategic planning, ultimately supporting concrete data monetisation.

## 4.2 Valuation Model

This section introduces the Valuation Model, which provides a relative value of a dataset by combining the metrics derived from the previous steps (the taxonomy process and the defined context). The resulting value does not represent a market price on its own. Rather, it serves as a comparative measure across datasets or as an input for benchmarking processes, provided that the same valuation strategy (including metrics and definitions) is consistently applied.

### 4.2.1 Valuation Models - Step I

Figure 3 shows the initial step within the valuation model. This step is only required in case the ANP process is not used. In other words, the strategy definition is not fed into Figure 2, and thus the context pass directly into the Valuation Model. This context is then used in Step I to define minimum considerations to perform for data valuation in Step II.

As observed in Figure 3, the first step involves a series of questions guiding the user towards a series of recommended indexes. These indexes are agglomerated in order to obtain recommendations to be used later on in the framework Step II and Step III (Figure 4 and Figure 5, respectively).

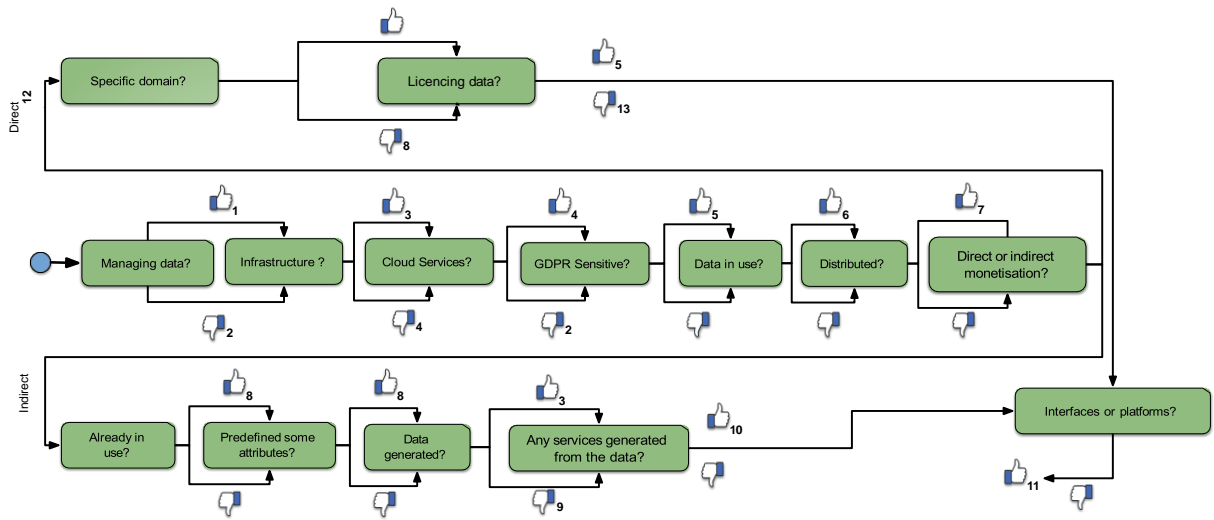

Figure 3: Valuation Model - Step I

For example, the first screening question "Does your service involve only managing and/or cleaning data?" assigns an index of 1 for "yes" and 2 for "no." Table 2 then summarises the checklist outcomes accumulated in Step I.

The primary goals of Step I are to decide (1) whether to pursue direct or indirect monetisation, (2) how relevant are expenditures and costs for data valuation, and (3) which core dimension - *Quality*, *Relevance*, or *Utility* - will drive the data valuation process. Understanding these dimensions and their relative importance is crucial for understanding the overall framework and its subsequent steps.

Table 1 provides a summary of the questions and their importance.

### 4.2.2 Valuation Model - Step II

Figure 4 presents the second step of the Valuation Model. As shown, users may enter this step either directly from the taxonomy tool or from the full evaluation performed in Step I.

In the latter case, users are guided through a series of questions to identify recommended metrics and KPIs for data valuation. Similarly to Step I, an indexing approach is used to define some evaluation characteristics that are passed in the last step. Table 3 summarise the results.

In order to determine which metrics are more relevant in this step, we rely on the classification used by author name [32]. Data can have four purposes:

| Question | Importance |
|---|---|
| Managing Data? | Establishes whether the entity only cleans/manages data or generates new data. If "yes," valuation should focus solely on cost-based components (Table 2 – Index 1). It distinguishes data-service providers (low added value) from data generators (higher potential for monetisation). |
| Infrastructure? | Determines if the organisation provides the technical infrastructure for data storage or management. A "yes" implies CAPEX + OPEX must be included in valuation (Index 3). It signals capital intensity and operational cost relevance. |
| Cloud Services? | Evaluates if cloud deployment costs are present, impacting OPEX in Step III (Index 4). Indicates recurring expenses tied to scalability, which directly affect the valuation's cost dimension. |
| GDPR Sensitive? | Identifies compliance and governance burdens (Index 5). If data involve personal or sensitive content, legal protection, anonymisation, and risk scores become mandatory valuation metrics. |
| Data in Use? | Reveals whether the dataset is already used elsewhere. If reused or benchmarked, quality and completeness metrics gain relevance; if unique, valuation must include relevance or utility assessment (Index 6). |
| Distributed? | Determines whether the data originates from multiple sources or nodes. Distributed datasets increase complexity and enhance an aggregation of metrics and costs should be performed (Index 7). |
| Direct or Indirect Monetisation? | Defines the strategic path: direct sale (e.g. DaaS) or internal exploitation. Affects both demand estimation and Information Cost Fraction (ICF). |
| Specific Domain? | Identifies sectoral specificity (e.g., energy, mobility, health). Narrow domains emphasise relevance and contextual fit over broad utility (Index 8). |
| Licencing Data? | Checks for ownership and usage rights (Index 5 and 13). Licensing directly impacts valuation since it governs revenue streams, renewability, and cost recovery terms. |
| Already in Use? | Detects whether the dataset is deployed in multiple environments, implying higher utility and demand potential (Index 8). If already exploited, valuation must incorporate multi-application benefit. |
| Predefined Some Attributes? | Indicates the dataset's maturity and structure. Predefined attributes implies also higher utility potential (Index 8). |
| Data Generated? | Determines whether data are actively produced (as opposed to passive management). Active generation implies inclusion of production costs (CAPEX) and potential exclusivity value (Index 3 and 5). |
| Any Services Generated from the Data? | Determines whether the dataset produces or supports downstream services (dashboards, APIs, analytics layers, recommendation engines, etc.). If "yes," the monetisation path aligns with Information-as-a-Service (IaaS) or Answer-as-a-Service (AaaS) models. These strategies require integrating demand estimation into the valuation (Table 2 – Index 10), since the data's value depends on the market or user consumption of the derived service rather than the data itself. In Step III, this translates to including Demand (and possibly price proxies) in the valuation equation |
| Interfaces or Platforms? | Identifies value-added layers (e.g., analytics, dashboards, APIs). If present, the valuation must include service-related metrics such as satisfaction or churn (Index 11). |

Table 1: Checklist of Dataset Valuation Recommendations

| Index | Recommendation (Results - Step I) |
|---|---|
| 1 | After Step I, if your data is normalised, go directly to Step III considering only Cost components. |
| 2 | Consider at least Quality (and the components of Step III) during dataset valuation (Step III). |
| 3 | Consider including OPEX and CAPEX during dataset valuation (Step III). |
| 4 | Consider including OPEX during dataset valuation including all the costs for (Step III). |
| 5 | Consider including data governance and legal compliance costs and metrics in Step III. Choose one or more: (1) Number of Sensitive Fields, (2) Privacy Level, (3) Risk Score, (4) Protection Expenses. |
| 6 | Consider evaluating the Relevance of the data, instead of Quality in Step III. Make sure to contain in the Relevance Evaluation each of the components defined for Quality Evaluation. |
| 7 | For each dataset, it is necessary to estimate its value based on their relative contributions (i.e. the features/variables/columns they bring in the global set). |
| 8 | Consider evaluating the Utility of the data, instead of Quality in Step III. Make sure to include in the Utility Evaluation each of the components defined for Quality Evaluation. |
| 9 | Consider all the costs related to Ownership and acquisition. Use metrics such as (1) Licensing, (2) Service Agreements, and (3) Value Added. |
| 10 | Consider the strategy as Information-as-a-Service or Answer-as-a-Service. Make sure to use Demand during dataset valuation (Step III). Use demand models or direct estimation. |
| 11 | Consider including different metrics related to quality of service during dataset valuation (Step III). Consider (1) Churn, (2) User Frequency, (3) Satisfaction, and/or (4) Reputation. |
| 12 | Consider your strategy as Data-as-a-Service and thus, use it for dataset valuation (Step III). Consider models such as Hotelling and CES or others for its estimation. |
| 13 | Since data will be sold and possibly lose ownership, consider this as a one time transaction. This imply in step III, use Information Cost Fraction ICF =1, independently of the uses. |

Table 2: Checklist Recommendations - Step I

| Index | Recommendation (Results – Step II) |
|---|---|
| 1 | Consider including Age, Timeliness, or any other time-dependent quality characteristics. Since data may serve multiple purposes, use the Information Cost Fraction (ICF) to estimate cost; carry this consideration into Step III. |
| 2 | Include Accuracy, Variety, Volume, and Completeness. If data is consumed only once, set the ICF to a one-time down payment of costs. |
| 3 | When no additional value arises after data collection, factor in privacy costs and governance/compliance overhead; in Step III, set Demand to 0, unless you have a clear representation of the demand for your service. |
| 4 | If data yields no immediate value but supports long-term analyses (ROI), apply other financial evaluation methods and set Demand to 0 in Step III, unless you have a clear representation of the demand for your service. |
| 5 | Add processing costs to the cost components in Step III. |
| 6 | Note that the taxonomy tool provides metric weights (relative importance), not their raw values. Users must evaluate and normalise each metric using the methods outlined here. |
| 7 | Consider including Accuracy and Completeness. Data can be used more than one time for generating different knowledge. For each DIFFERENT possible applications, consider an extra value for Information Cost Fraction (ICF) (e.g. if there is an application for process optimisation of the data set and another application for data trading, ICF = 2. |

Table 3: Checklist Recommendations – Step II

- *Operational Data* - data that is used on a regular basis by an organisation at a frequency determined by its application.

- *One Time Decision* - data that is used for infrequent and often one-off decisions within an organisation.

- *Legal and Safety* - This data often yields no value to an organisation other than regulatory and legal costs or prevention of injury and damages.

- *Research and Innovation* - This data has the innate attribute of being high risk high reward.

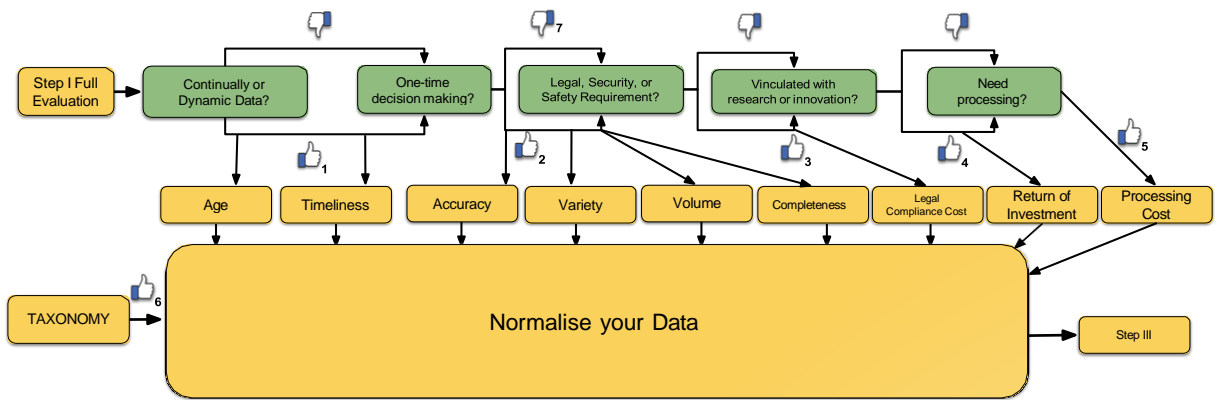

Figure 4: Valuation Model - Step II

Independently of the approach used to define metrics and KPIs, Step II also involves normalising the selected (and measured) metrics and KPIs. It is important to note that these recommendations represent only a minimum baseline. Users are strongly encouraged to incorporate additional metrics and KPIs wherever possible to strengthen the valuation.

As observed in Figure 4, after defining a set $M = m_i, i \in 1 \ldots n$ of metrics and KPIs, each metric is normalised ($m_{norm}$). As a recommendation, the normalisation can follow three approaches depending on the nature of the metric. This is detailed in subsection 4.3.

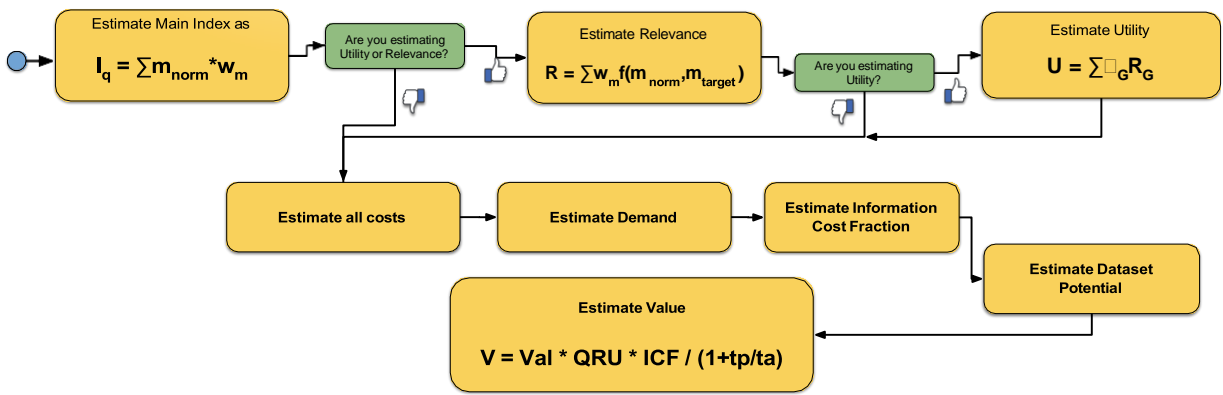

Figure 5: Valuation Model: Step III

### 4.2.3 Valuation Model - Step III

Finally, Figure 5 illustrates the schematic representation of the final step in the framework. This step integrates the context and metrics defined in Step I and Step II to carry out the data valuation. As shown in the figure, the process begins with the estimation of a main index. This index captures the combined contribution of previously normalised metrics that together reflect a single underlying concept (e.g. data quality). The index is computed as:

$$I_q = \sum_x m_{norm} \times w_m \tag{1}$$

where $w_m$ represents the weight assigned to each metric in the calculation of the index $I_q$. The summation covers all metrics $m$ that define the underlying concept $q$. These weights can be obtained through decision-making processes, assuming that $\sum w_m = 1$ and setting values, or, as in the use cases of this project, by employing ANP-based tools that provides specific $w_m$ values. The resulting index $I_q$ may correspond directly to a key performance indicator (KPI), providing a clear measure of strategic progress, or it may act as an aggregated metric in distributed systems, where $w_m$ reflects each component's relative contribution. Importantly, this agglomeration is optional and is only applied when users choose to evaluate higher-level indexes.

In cases where the Relevance $R$ is to be estimated, based on the outcomes of Step I, Step II, or the relative importance of the context, for each of the $G$ areas, domains, or applications under evaluation, it is computed using the following expression (as illustrated in Figure 5):

$$R_G = \sum_m w_m\, f(m_{norm}, m_{target}) \quad \forall\ G \tag{2}$$

The function $f(m_{norm}, m_{target})$ captures how closely the observed metric $m_{norm}$ aligns with its strategic target $m_{target}$. Note that $I_q$ can also serve as a proxy for $m_{norm}$ (i.e., an overall quality index), which will require an overall general target (i.e. an overall strategic quality target). The parameter $m_{target}$ represents limits or goals defined by enterprise strategies, which are not constrained by natural bounds such as $m_{max}$ or $m_{min}$ (defined in Section 4.3).

$$f(m_{norm}, m_{target}) = \begin{cases} \dfrac{m_{norm}}{m_{target}}, & if\ m_{norm} \geq 0,\ m_{target} > 0, \\ \dfrac{m_{target}}{m_{norm}}, & if\ m_{norm} < 0,\ m_{target} < 0, \\ \dfrac{m_{target}}{|m_{norm}|}, & if\ m_{norm} < 0,\ m_{target} > 0, \\ m_{norm}, & if\ m_{target} = 0, \end{cases} \tag{3}$$

Although a simple ratio is often used for efficiency-type metrics (so that $f = 1$ when $m_{norm} = m_{target}$ and falls toward zero as the gap widens), other forms may be more appropriate depending on the decision context.

$$f(m_{norm}, m_{target}) = \frac{1}{1 + m_{norm} - m_{target}} \tag{4}$$

A reciprocal formulation, smoothly penalises deviations by mapping any difference to a value strictly between zero and one, but penalise deviation in any direction from the target. This type of constraint is useful for cases requiring to match exactly the target (e.g. rack temperature within a data centre). For cases where steep penalties for even small deviations are desired-or where the distance should decay exponentially-one can instead use

$$f(m_{norm}, m_{target}) = \exp(-|m_{norm} - m_{target}|). \tag{5}$$

All variants share the property that $f$ reaches a value of one when the observed metric matches its strategic goal, decreasing monotonically toward a lower bound (often zero) as the discrepancy increases. The choice among them should depend on the metric's nature and the decision-maker's tolerance for deviations from the benchmark.

Once each domain's $R_G$ is obtained, the dataset's overall Utility $U$ is calculated as the sum of these values weighted by the relative importance $\beta_G$ of each domain with respect to the overall strategy. Unlike previous weightings, $\sum \beta_G = 1$ because these coefficients reflect contributions to an absolute total value; thus, the more applications across different domains, the greater the dataset's overall worth.

$$U = \sum_{G=1}^{G_{max}} \beta_G R_G \tag{6}$$

In the final stage of the valuation framework, after computing either the aggregated Quality, Relevance, or Utility index, the model requires three additional estimations (depending on Step I feedback): total costs, demand projections, and the Information Cost Fraction. The first two are used to estimate an intrinsic Dataset Potential (Val), before arriving at the dataset value $V$.

First, all relevant costs are aggregated. These include capital expenditures (CAPEX) for hardware and infrastructure deployment (over the analysis period, which may encompass processing time $t_p$ and acquisition time $t_a$), operating expenditures (OPEX) for maintenance, energy, storage, and processing, and any data-governance or legal-compliance expenses identified in Step I. Denoting the total outlay as Costs, this step ensures the valuation fully reflects the resources required to generate and deliver the dataset.

Second, demand for the service must be estimated. When available, market forecasts, historical sales data, or direct customer commitments can be used to project both a demand quantity *Demand* and an associated unit price $Price_{set}$. If such information is unavailable, an expected maximum gain $V_{max}$ (for example, based on expert judgment of the upper revenue bound for perfect data) and a minimum gain $V_{min}$ can be elicited and used as described later.

Once these elements are in place, the Dataset Potential Value *Val* [32] is determined by one of three approaches, applied in sequence depending on the level of information acquired:

$$Val = \begin{cases} Demand \times Price_{proxy/set} - Costs, & \text{if Demand, Price proxy/set, and Costs are known} \\ \frac{V_{max}+V_{min}}{2}, & \text{if only bounds } V_{max} \text{ and } V_{min} \text{ are available} \\ & \text{otherwise.} \end{cases} \tag{7}$$

Importantly, once a Dataset Potential Value is defined, and if comparison is the primary goal, the same expression from the referenced equation should be applied. The $price_{proxy/set}$ reflects the value of the information or service produced from the acquired data and should therefore be treated separately from costs. In distributed systems, the cost calculation must account for each component's relative contribution, so *Val* should represent the sum of all component costs. Since components may be subject to different acquisition contracts, each can have its own Information Cost Fraction for the period of analysis. Finally, in case that no price estimation and cost are involved (i.e. Val=1), the dataset value would considered the derived indexes as main proxy for value estimation (i.e. the value will scale only from Quality, Relevance, or Utility, not from direct economical factors).

Third, the Information Cost Fraction (*ICF*) is selected based on the monetisation strategy. A one-time sale typically sets $ICF = 1$, while subscription- or service-based models may use $ICF < 1$ to reflect retained rights or repeated usage. In some cases, *ICF* can exceed 1 to capture multi-utility scenarios; for example, in indirect monetisation where the same data supports multiple applications. This adjustment is only applied if Utility is not the primary metric in the valuation, since Utility inherently accounts for multiple scenarios.

Finally, the dataset Value *V* is computed using a modified formulation from [32]. This estimation combines the intrinsic potential Val, the total Main Valuation Driver *QRU*, the Information Cost Fraction ICF, and a temporal adjustment for processing time/delay $t_p$ relative to acquisition time $t_a$:

$$V = \frac{QRU \times \sum(\text{Val} \times \text{ICF})}{V_P} \qquad (8)$$

In this expression *QRU* represent one component between Quality *Q*, Relevance *R*, or Utility *U*, depending on the level of analysis defined by the framework user.

In [32] formulation the value component is represented as $V = V_N \times V_R / V_p$, where $V_n$ represented the multiplication of *Val* and a quality indicator, leaving outside higher perspectives as Relevance and Utility. Furthermore, in our work, *Val* contain an extended expression that consider demand and costs. Additionally, as observed in our estimation, a summation is used to represent distributed systems, a concept not cover in other works. $V_R$ represents the percentage contribution of a given data input to the overall value of its Decision Node. To foster its incorporation for direct and indirect monetisation, its definition was covered by a modified version of *Val* and the incorporation of the *ICF*.

As described in [32], the corrective factor $V_p$ adjusts the estimated value according to acquisition and processing times. Depending on the type of data, it takes one of two forms: For data with significant processing time $t_p$, $V_p = 1 + T_p/T_a$. For data that is purchased or requires negligible processing (as defined by [32]), a fixed contribution is applied: $V_p = 1 + 0.05$ down to $V_p = 1 + 0.01$. Although the equation implies a reduced price as processing time increases, its impact differs for buyers and providers. Buyers see additional processing as a value loss, while providers associate it with higher costs, potentially lowering both the intrinsic value (Val) and the corrective factor. Between two datasets offering equivalent information, the one requiring more processing is inherently less valuable. Since $V_p$ is a relative measure, it can still be used to compare datasets as long as the same formulation is applied consistently.

Together, these expressions complete the Step III valuation framework, linking normalised quality and cost measures through relevance or utility to produce a defensible price or score. By estimating comprehensive costs, forecasting or bounding demand, selecting an appropriate information fraction, and computing dataset potential, organisations can translate abstract quality and utility measures into valuations that align technical performance with market realities.

### 4.3 Normalisation

These can be linear, delimited or fixed, and binary. In the linear normalisation, the metric is linearised between maximum ($m_{max}$) and minimum ($m_{min}$) values that do not necessarily reflect the desired strategic values; it can reflect natural limits within the metric - e.g. maximum memory and total number of dataset columns. By doing so, an indicator is constructed $m_N$ for the metric $m$. The linearisation, $m_N = \xi_m \times \frac{m - m_{min}}{m_{max} - m_{min}}$ contain a weight value for the metric $m$ that represent the impact of the metric in the overall dataset value. $\xi_m$ is used to correct the representation of inversely correlated metrics with respect to the dataset value. $\xi_m$ can assume a value of 1 for metrics that are positively correlated and -1 otherwise.

Delimited normalisation assigns a value of 1 when the metric exceeds a threshold $m_t$; otherwise it is 0. The threshold depends on the metric or domain and is not necessarily tied to strategy. Finally, binaries transform categorical variables with two possible values into 0 or 1. If several categories do exist, a range of values could be defined depending on the hierarchical importance of the categoires within the index.

## 5 Detailed Examples and Validation

### 5.1 Example I - Selling Data

Here's an imaginary case for GreenRoute Analytics, which deploys roadside sensor arrays to capture real-time traffic flow, air-quality and noise-level data for smart-city planners. They both collect and preprocess the feeds, host them on proprietary edge-and-core infrastructure, and then package the cleaned streams for direct sale to municipal partners (i.e. DaaS). The main purpose of GreenRoute Analytics is not to internally use the generated data and thus the loss of data ownership is not contradictory with their main strategy. Furthermore, even though they have several clients, the particular dataset considered by the municipal partner has no further use by other stakeholders and thus, the costs involved into its acquisition and maintenance are exclusive for this partner.

Since GreenRoute is in a rush to perform a valuation of the dataset, they are interested into using a simplistic approach to evaluate the relative value of their data to offer it to the municipal partners. To assist their valuation process, they have a history of selling similar datasets to the same municipal partner.

#### 5.1.1 Example I - Step I

As outlined in the framework, the first stage of the valuation process is to establish the context and assess how it will influence the evaluation approach. In this case, GreenRoute is not a dedicated data provider and seeks a simplified assessment of its dataset. Consequently, it applies the decision tree from Step I rather than the taxonomy-based method described in [9] . Table 4 presents the evaluation process for Step I, including the connector codes.

As shown in Table 4, the evaluation yields Connector Codes 2, 3, 12, and 13. These correspond to: (2) considering at least data quality; (3) including both OPEX and CAPEX in the dataset valuation; (12)

| Step | Question | Answer | CC |
|------|----------|--------|-----|
| 1 | Does your service involve ONLY managing and/or clearing data? | No - they collect raw sensor streams and also apply real-time cleansing and aggregation. | 2 |
| 2 | Are you providing infrastructure for managing and/or containing the data? | Yes - they operate edge gateways at lamp-post units and a private data-hub backend. | 3 |
| 3 | Are you using cloud services to keep data? | No - all storage and processing runs on their own hybrid on-premise clusters. | 2 |
| 4 | Is the data GDPR-sensitive or sensitive to any legal consideration? | No - measurements are environmental only, with no personal identifiers. | none |
| 5 | Is the data already in use in other domains/areas? | Yes - similar traffic and pollution feeds are sold to logistics and urban-planning firms. | 6 |
| 6 | Is the data already in use in other domains/areas? | No - the precise combination of noise, air-quality, and vehicle-counts is unique. | none |
| 7 | The data is for direct monetisation (selling) or indirect monetisation (competitive advantage)? | Direct - GreenRoute sells normalised data packages to city authorities. | 12 |
| 8 | Is the data used by an exclusive number of users, or a specific domain? | Yes - only accredited municipal agencies and their consultants may access it. | none |
| 9 | Are you licensing your data? | No - they offer one-year subscriptions rather than perpetual licenses. | 13 |
| 10 | Are you providing interfaces and platforms for clients? | No - data is delivered via nightly batch exports, not via web/UI. | none |

Table 4: GreenRoute Analytics: Step I Pipeline Responses with Connector Codes (CC)

framing the strategy as Data-as-a-Service and considering demand models such as Hotelling and CES; and (13) treating the sale as a one-time transaction with potential loss of ownership, using an Information Cost Fraction $ICF = 1$.

From a hierarchical perspective, the results indicate that estimating dataset Quality is a necessary metric to track, even though higher-level indexes such as Relevance or Utility could also be used. At a minimum, the valuation model must incorporate robust quality metrics, such as accuracy, completeness, and timeliness. These minimum quality considerations are addressed in Step II, though the scope of metrics can be extended beyond the predefined set as needed.

In parallel, both capital expenditures (CAPEX) associated with sensor deployment and edge-gateway hardware, and operating expenditures (OPEX) covering maintenance, power, and network services could be fully accounted for cost base estimation. Beyond raw quality, the valuation framework should evaluate the relevance of each metric (e.g. feature-noise levels, vehicle counts, air-quality indexes) by embedding standard quality components within a relevance assessment that reflects the relative contribution of each variable to the end user's objectives.

Strategically, GreenRoute should treat its offering as a one-off data sale rather than a Data-as-a-Service model, since the dataset is delivered once and ownership is transferred to the buyer. Accordingly, the Information Cost Fraction (ICF) is set to 1, reflecting a single transaction and simplifying revenue recognition.

After completing Step I, GreenRoute Analytics must proceed to a comprehensive valuation of its dataset in Step II. Although the data is already cleaned and some metrics are already measured, it has to be normalised for agglomeration.

### 5.1.2 Example I - Step II

In the second stage of the Data Monetisation Framework, GreenRoute Analytics confirms that its roadside sensor data is continually operational, used for a one-time decision process, free from legal or safety requirements, unrelated to research or innovation grants, and requires cleansing and aggregation to extract value. The framework therefore prescribes tracking accuracy, variety, volume, and completeness, along with processing costs. While industry best practices recommend monitoring data age and timeliness, GreenRoute omits these since the data is generated and sold immediately upon collection, and the unique metric combination makes such measures unnecessary. Legal compliance costs and ROI metrics are likewise excluded,there are no regulatory burdens, and the one-off sale with already contractual terms does renders ROI tracking irrelevant.

To normalise the selected quality and cost metrics (accuracy, volume, processing costs, and completeness), both an observed value and an expected benchmark are defined for each:

Accuracy: observed at 0.80 (scale 0–1), benchmark 1.00.

Volume: 12,785,568 records during the period, benchmark 20,000,000 records.

Processing cost: $887 per terabyte, benchmark (maximum acceptable) $200 per terabyte.

Completeness: 400 empty records out of 12,785,568, benchmark maximum 500 empty records out of 20,000,000.

Linear contribution scores ($I$) for continuous measures are then computed as the absolute difference between the observed value and the lower bound (zero or zero cost), divided by the span between the benchmark and the lower bound, with unity weight. For accuracy, for example:

$$Accuracy_{norm} = \frac{|0.80 - 0| \times 1}{|1.00 - 0|} = 0.80.$$

For volume:

$$Volume_{norm} = \frac{|12,785,568 - 0| \times 1}{|20,000,000 - 0|} \approx 0.6393.$$

For processing cost, because higher cost diminishes value (i.e. inversely correlated to data value), we treat the $\xi_m$ term as -1 and compare the observed cost to the minimum acceptable:

$$PCost_{norm} = \frac{|887 - 0| \times (-1)}{|200 - 0|} \approx -4.435.$$

This negative score reflects that the observed cost is more than four times the expected maximum, signaling a substantial drag on dataset value from an operational-cost perspective.

Completeness, being categorical (a fixed proportion of empty versus non-empty entries), is handled by a fixed-contribution rule. We compare the observed emptiness ratio (400/12,785,568 $\approx 3.129 10^{-5}$) with the maximum allowable emptiness ratio (500/20,000,000 $= 2.500 10^{-5}$). Because the observed ratio exceeds the expected threshold, the completeness contribution is set to one ($m_{Completeness} = 1$).

Through this normalisation process, GreenRoute transforms heterogeneous real-world observations into comparable contribution scores. These scores then feed directly into the valuation model, ensuring that accuracy, dataset size, cost efficiency, and data completeness each factor in proportionally to their normalied contributions.

In addition to accuracy, volume, processing cost, and completeness, the Data Monetiation Framework recommends evaluating data variety, the diversity of sources, formats, or feature types, and usage metrics such as user access frequency. For GreenRoute, tracking variety was infeasible because the raw sensor feeds consist solely of homogeneous environmental telemetry, and the planned one-off batch delivery model made user frequency measurement irrelevant. Consequently, variety and access metrics were excluded from normaliation and valuation, as the transaction involves a direct data sale without ongoing access services. For future iterations, the data pipeline could be enriched with additional sensor modalities, for example, integrating video-based information alongside existing sensor data.

### 5.1.3 Example I - Step III

In the final valuation stage, the normalied contribution scores are aggregated into a single quality index. Using the previously computed values, accuracy at 0.80, volume at 0.6393, processing cost at –4.435 (indicating cost drag), and completeness at 1.00, each quality dimension is weighted equally ($z = 1/3$ for accuracy, volume, and completeness - Each quality dimension was waighted equally since no full ANP process was performed to define their relative importance).

Based on the Step I outcome, processing cost is excluded here since the primary quantification metric is Quality. The quality index is therefore calculated as:

$$I_q = \tfrac{1}{3} 0.80 + \tfrac{1}{3} 0.6393 + \tfrac{1}{3} 1.00 \approx 0.8131,$$

which serves as our proxy for overall dataset "goodness" when demand and direct cost information are unavailable.

Although GreenRoute's strategic focus is on evaluating data Quality, rather than computing an aggregate of Relevance or Utility metric, we nonetheless illustrate the full Utility-based valuation here, as a concrete example across two distinct user domains, each characterized by its own reference benchmarks.

Beginning with the previously derived quality index $Quality \approx 0.8131$, we compute domain-specific relevance scores $R$ via a balanced normaliation against each domain's respective reference values $Quality_{target}$ and $PCost_{target}$. Furthermore, we settle that each index has similar relative importance (which could be extracted using the taxonomy tool [11], thus $w_m = 0.5$ for Quality and Processing Cost Indexes. In Domain 1, where $Quality_{target} = 1.00$ and $PCost_{target} = -2.5$, the relevance score is

$$R_1 = 0.5 \frac{0.8131}{1.00} + 0.5 \frac{-2.5}{-4.435} \approx 0.6884.$$

In Domain 2, with less stringent benchmarks $Quality_{target} = 0.95$ and $PCost_{target} = -3.0$, the calculation yields

$$R_2 = 0.5 \frac{0.8131}{0.95} + 0.5 \frac{-3.0}{-4.435} \approx 0.7662.$$

As expected, the relative value of the dataset would be higher in the less restrictive context. To estimate the utility, instead of the relevance, the relevance are summed up. As pondering factor we use $\beta_G = 1$ since both domains represent similar market opportunities. Summing these gives a total utility $U = \beta_{G,1} R_1 + \beta_{G,2} R_2 \approx 1.455$.

Finally, using the Information Cost Fraction (ICF) determined in Step I, and noting that the data is not distributed, as it is generated centrally, we need only calculate $V_P$ and the intrinsic potential (*Val*) to complete the valuation.

First, as recommended in Step I, we calculate both OPEX and CAPEX for the valuation. For CAPEX, since the sensors are used exclusively for this data generation, they can be treated as a depreciated asset. Assuming a sensor cost of $5,000 with a lifespan of five years, and data generated over a single month, the monthly depreciation is: $\$5,000/5[year] \times 12[month/year] \approx \$83.34[1/month]$. For OPEX, costs include the previously specified processing cost ($887 / TB) plus other operational expenses estimated at $500 / TB. Given that the dataset volume is 0.6 TB (as measured by GreenRoute), operational costs are: $(\$887[1/TB] + \$500[1/TB]) \times 0.6[TB/month] = \$832.2/month$. Then the total costs involved for the GreenRoute dataset can be estimated as $\$832.2[1/month] + \$83.34[1/month] = \$915.54[1/month]$. Since demand is fixed but the ideal price is not yet set, it can be left as a variable or estimated by assuming a fixed profit margin, for example, 10 %. In that case: $Val = Demand \times Price - Costs = \$915.54 \times 0.1 = \$91.5$.

For the temporal adjustment, because the processing time is significant relative to the acquisition time, we apply $V_P = 1 + T_p/T_a$. With a processing time of 15 days over a 30 day sensing period, the final valuation is calculated as follows:

$$V = \frac{Val \times U \times ICF}{1 + \frac{t_p}{t_a}} = \frac{\$91.5 \times 1.455 \times 1}{1.5} \approx \$88.8.$$

This final value estimation is relative; therefore, similar datasets can be compared provided that their utility can be estimated under comparable assumptions (i.e., similar *ICF*, $t_p$, and $t_a$). This exercise illustrates that, even when prioritiing relevance, the utility framework can be applied to quantify value across heterogeneous domains by adapting each domain's reference parameters.

## 5.2 Example II - Buying Data

This case study examines a data-buying scenario from the perspective of FlowSmart, a European startup specialising in micro-mobility services. To optimise its e-scooter deployment strategy across five major metropolitan areas, the company identified a need for high-resolution geolocation data; specifically, timestamped, anonymised user location traces collected at five-minute intervals over the past twelve months. Three third-party providers, including commercial data brokers and mobile application vendors, offered datasets meeting the general specifications. In some cases, datasets would need to be merged to cover all five areas (i.e., one vendor offers full coverage, while the other two provide complementary subsets).

To guide the selection process, FlowSmart's data science and procurement teams applied our structured data valuation framework with the objective of comparing competing datasets across economic, quality, contextual, and governance dimensions, ensuring a transparent and evidence-based decision. While all providers claimed similar geographic and temporal coverage, differences emerged in pricing models, licensing restrictions, and data quality guarantees.

FlowSmart estimated that integrating high-quality geolocation data could improve fleet utilisation by approximately 20% through enhanced demand prediction and real-time rebalancing algorithms, corresponding to a projected annual revenue gain of up to $80K. However, with only partial visibility into the datasets prior to purchase, the company adopted a conservative forecast. Acquisition costs, excluding post-processing and compliance assessments, were set by the vendors. To avoid post-processing, FlowSmart required that datasets be delivered in CSV format with a precision of three decimal points.

### 5.2.1 Example II - Step I

For this analysis, FlowSmart opted to use the taxonomy tool [11] rather than a direct evaluation via the decision tree. By combining the tool with a full understanding of their operational context, the company identified the most relevant metrics for tracking the **Financial**, **Data Quality**, **Context**, and **Governance & Compliance** dimensions.

As shown in Figure 1, each of these dimensions can be mapped to data clusters within the **Financial** and **Internal Process** perspectives of the Balanced Scorecard (BSC). Using the taxonomy [9] and the tool, the following metrics were selected along with their relative importance:

Table 5: Assessment for Candidate Datasets

| Dimension | Metric[a] | Description | $w_m$[β] | D1[γ] | D2[γ] | D3[γ] |
|---|---|---|---|---|---|---|
| Financial | Risk Cost (RC) | Account for the costs arising from loss, compromise, or misuse of data. | 0.06 | $0 | $2K ×2 | $1K×2 |
| Financial | Storage Cost (SC) | Expenses of storing datasets | 0.1 | $10K | $7K | $7K |
| Financial | RoI | Evaluating these costs relative to financial returns | 0.3 | 70k - COSTS | | |
| Quality (Q) | Granularity (G) | Detail or precision at which data are captured | 0.2 | 1 [min] | 5 [min] | 10 [min] |
| Quality (Q) | Format (F) | Define both the structure of the data and the percentage of compliant | 0.03 | Json | CSV | CSV |
| Quality (Q) | Precision (P) | Define both the structure of the data and the percentage of compliance to it | 0.03 | 3 decimal | 2 decimal | 3 decimal |
| Quality (Q) | Timeliness (T) | Time between an event and the data being available to the business | 0.05 | 1.5 [year] | 1 [year] | 1 [year] |
| Context | Relevance | Usefulness of data for a given context | - | To estimate | | |
| Quality | Similarity (S) | Evaluate similarities or differences between sets, vectors and properties. | 0.2 | 4/5 | 3/5 | 5/5 |
| Governance | Compliance (Com) | Track adherence to legal, standards, and regulatory requirements. | 0.03 | yes | no | no |
| Financial | price | selling price | NA | 15K | 10K | 10K |
| Financial | Processing Cost (PC) | Not considered as a metric but reported for detailed information | NA | $0K | $2K | $1K |

[a] The metrics mentioned here are defined by using the taxonomy tool.
[β] $w_m$ represent the relative importance to estimate the Relevance of the overall dataset.
[γ] Represent the value of the metric from the different sources of data (i.e. distributed system).

Table 5 lists the metrics and KPIs provided by each vendor (*D*) that are relevant for the dataset comparison. **Relevance** is not included in the estimates, as it serves solely to define hierarchical weighting within the valuation framework (i.e., it represents an aggregation of indexes pertinent to the context).

Among the metrics, **Similarity** is defined by the user as the degree of overlap between a vendor's coverage and the target metropolitan areas. In this case, the first vendor (D1) covers four out of the five areas but can be combined with the second vendor's (D2) data to achieve full coverage.

The **Governance & Compliance** metric specifies whether the data is already anonymised or requires further processing, which would entail additional costs. FlowSmart estimated these costs at $0K for D1, $2K for D2, and $1K for D3. These anonymisation costs also serve as the basis for calculating the **risk cost**, defined here as twice the anonymisation expense incurred by the company (see Table 5).

### 5.2.2 Example II - Step II

As defined in the framework, the first step is the **normalisation** of metrics to obtain suitable indexes. Based on the following user-specified limits the normalisation is performed:

Granularity: 1 minute; Format (CSV): 1 if CSV, 0 otherwise; Precision: 1 if 3 decimals, 0 otherwise; timeliness: 1 year; Similarity: maximum value = 5/5; Compliance: 1 if anonymised, 0 otherwise; Risk Cost: ideal = $0; Storage Cost: ideal = $5K; RoI: ideal = $80K

To normalise the selected metrics, a linear or binary normalisation is applied. The following normalisations are used:

**Granularity (G)** - positively correlated; higher granularity is better, but given its units for representation, the lower the better. Therfore, we can transform its representation as data points per hour, facilitating it to be positively correlated, then assuming $m_{max} = 60[data\ points/hour]$, $m_{min} = 6[data\ points/hour]$: $G(D_1)_N = \frac{60-6}{60-6} = 1.0$; $G(D_2)_N = \frac{12-6}{60-6} = 0.11$; $G(D_3)_N = \frac{6-6}{60-6} = 0.0$.

**Format (F)** - We assume a binary evaluatoin for CSV format as positive, then: $F(D_1)_N = 0$ (JSON); $F(D_2)_N = 1$ (CSV); $F(D_3)_N = 1$ (CSV).

**Precision (P)** - we assume a binary evaluation for precision as if the dataset fullfills or not the requirement: $P_{D1} = 1$ (3 decimals); $P_{D2} = 0$ (2 decimals); $P_{D3} = 1$ (3 decimals)

**Timeliness (T)** - For our case, the higher the timeliness the better since we consider it correlated to the time start for aquisition. The combination of it with the granularity/frequency of aquisition can be used to estimate the data volume too. Given that it is possitively correlated and assuming $m_{min} = 0$, $m_{max} = 1.5$: $T(D_1)_N = \frac{(1.5-0)}{1.5} = 1.0$; $T(D_2)_N = \frac{1-0}{1.5} = 0.67$; $T(D_3)_N = \frac{1-0}{1.5} = 0.67$

**Similarity (S)**: $S(D_1)_N = \frac{4}{5} = 0.8$; $S(D_2)_N = \frac{3}{5} = 0.6$; $S(D_3)_N = \frac{5}{5} = 1.0$

**Compliance (Com)** - we consider compliance with the anonymised requirement as binary, then: $Com(D_1)_N = 1$ yes; $Com(D_2)_N = 0$ no; $Com(D_3)_N = 0$ no.

**Risk Cost (RC)** - Negatively correlated Assume: $m_{min} = 0$, $m_{max} = 2000$
$RC(D1)_N = \frac{0-0}{2000}(-1) = 0.0$; $RC(D_2)_N = \frac{2000-0}{2000}(-1) = -1.0$; $RC(D_1 + D_2)_N = RC_1 + RC_2 = -1.0$; $RC(D_3)_N = \frac{1000-0}{2000}(-1) = -0.5$.

To facilitate the calculation and presentation of the example, the aggregated costs needed to represent the overall dataset (given the fractioning of the information between $D_1$ and $D_2$), are included here and in the following costs.

**Storage Cost (SC)** - Negatively correlated and assuming $m_{min} = 5K$, $m_{max} = 15K$: $SC(D1)_N = \frac{10K-5K}{10K}(-1) = -0.5$; $SC(D_2)_N = \frac{7K-5K}{10K}(-1) = -0.2$; $SC(D_1 + D_2)_N = \frac{10K+7K-5K}{10K}(-1) = -1.2$; $SC(D_3)_N = \frac{7K-5K}{5K}(-1) = -0.2$.

**Return on Investment** - Positively correlated and assuming a desired return on investment in percentage we can use the cost as maximum referencing value. Furthermore, a fraction of their contribution (based on Similarity) can be used to estimate revenue. Then, each RoI proxy can be calculated as $\frac{70K-Costs}{Costs}$. Then:

- $RoI(D_1)_N = \frac{70K*4/5-(0K+10K+0K+15K)}{(0K+10K+0K+15K)} = 1.24$
- $RoI(D_2)_N = \frac{70K*3/5-(4K+7K+2K+10K)}{(4K+7K+2K+10K)} = 0.82$
- $RoI(D_1+D_2)_N = \frac{70K-(0K+7K+0K+15K)-(4K+7K+2K+10K)}{(0K+7K+0K+15K)+(4K+7K+2K+10K)} = 0.56$
- $RoI(D_3)_N = \frac{70K-(2K+7K+1K+10K)}{(2K+7K+1K+10K)} = 2.5$

As observed in the estimation of *RoI*, the aggregated index ($D_1+D_2$) is lower than both individual components. This is explained by the summation of both costs that overlap some of the information provided.

### 5.2.3 Example III - Step III

In the final stage of the data-buying scenario, FlowSmart conducts the valuation of each candidate dataset (D1, D2, and D3) as well as the combination of D1 and D2, using the normalised metrics obtained in Step II. The objective is to estimate the relative value of each dataset (and their combination) by applying the valuation model.

Since multiple metrics fall within the quality dimension, the index formulation can be used to aggregate them into a single global quality metric, even if this metric is not directly used in the estimation of relevance. This aggregated value can be computed by weighting the contributions of each dataset according to their relative importance, producing an overall quality measure. Then:

$$I_{q,norm} = 0.2 \times G_N + 0.03 \times F_N + 0.03 \times P_N + 0.05 \times T_N + 0.2 \times S_N \tag{9}$$

Then, it is obtained that $I_{q(D_1),norm} = 0.44$, $I_{q(D_2),norm} = 0.21$, and $I_{q(D_3),norm} = 0.29$. Using an assumption that each dataset fraction quality index contributes to the aggregated dataset, the aggregated data quality index $I_{Q(D_1+D_2),norm}$ can be estimated as:

$$I_{q(D_1+D_2),norm} = \frac{4 \times I_{q(D_1),norm}}{5} + \frac{3 \times I_{q(D_2),norm}}{5} = 0.48. \tag{10}$$

Since the fractions were not imposed over all the indexes before hand, the quality indexes could be furthermore constrained in order to have a wider representation of the dataset. Furthermore, some synergetic interactions could be consider when merging both datasets, nevertheless, since a full knowledge of the dataset is not given in this exemplification, we use only the fraction components instead of analyzing the full aggregated dataset. Furthermore $w_{quality}$ can be estimated as the aggregation of the different contributions ($w_{quality}$ = 0.2 +0.03+0.03+0.05+0.2 = 0.51). It is important to consider that quality, as expressed right now, each of their components is positively correlated and thus they do not contradicts the estimation of $f(m_N, m_{target})$ expressed as relative efficiency.

As defined, relevance will be used as the main component to estimate the values of the dataset. Then, following its expression, we can calculate it as (assuming only one G domain and the representation of $f(m_N, m_{target})$):

$$R_G = \sum_m w_m \, f \, m_N, m_{target} \quad \forall \, G =$$
$$w_Q f(Q_N, Q_{target}) + w_{RC} f(RC_N, RC_{target}) + w_{RC} f(SC_N, SC_{target}) +$$
$$w_{RoI} f(RoI_N, RoI_{target}) + w_{Com} f(Com_N, Com_{target})$$

Assuming that the strategy is to maximize the benefits and favorable conditions of the datasets, the targets for each metric are set to their optimal values (e.g., $Q_{target} = 1$ and $RC_{target} = 0$). By setting the cost target to zero, the function $f$ reduces to the metric itself. Then:

$$R_{D1} = 0.51 \times \frac{0.44}{1} + 0.06 \times 0 + 0.1 \times -0.5 + 0.3 \times \frac{1.24}{1} + 0.03 \times \frac{1}{1} = 0.576 \tag{11}$$

$$R_{D2} = 0.51 \times \frac{0.21}{1} + 0.06 \times -1 + 0.1 \times -0.2 + 0.3 \times \frac{0.82}{1} + 0.03 \times 0 = 0.273 \tag{12}$$

$$R_{D3} = 0.51 \times \frac{0.29}{1} + 0.06 \times -1 + 0.1 \times -0.2 + 0.3 \times \frac{2.5}{1} + 0.03 \times 0 = 0.818 \tag{13}$$

$$R_{D_1+D_2} = 0.51 \times \frac{0.48}{1} + 0.06 \times -1 - 0.12 + 0.3 \times \frac{0.56}{1} + 0.03 \times 0 = 0.233 \tag{14}$$

Although the results might initially suggest that combining datasets increases overall relevance (i.e. $R_{D_1} + R_{D_2} > R_{D_3}$), the aggregated option also incurs substantially higher costs. Therefore, the combined dataset's relevance must be recalculated as a single aggregated metric, or alternatively, the final decision should be based on the complete value estimation.

The value of each dataset (and the combination) is calculated using relevance and the primary contributing factor. Since processing time and acquisition time are not known or negligible, a constant factor, as defined by [32] is used for $V_P$ (i.e., $V_P = 1.05$). As the transaction is a one-time acquisition, the Information Cost Fraction is set to unity ($ICF = 1$). The valuation ($Val$) is then estimated for both best- and worst-case scenarios using $Val = (V_{max} + V_{min})/2$ where $V_{max}$ corresponds to the full profit scenario ($V_{max} = RoI$) and $V_{min}$ represents the cost-only case ($V_{min} = -\text{Costs}$). Then:

$$V_{D_i} = \frac{Val_{D_i} \times R_{D_i} \times ICF}{V_p} = \frac{\frac{RoI-RC-SC-PC-Price}{2} \times R_{D_i}}{1.05}. \tag{15}$$

Importantly, the estimation of Val is not an index and thus it should preserve the same units as *price*. Then

$$V_{D_1} = \frac{\frac{70K \times 4/5 - (0K+10K+0K+15K) - 0K - 10K - 0K - 15K}{2} \times 0.576}{1.05} = 1.65 \tag{16}$$

$$V_{D_2} = \frac{\frac{70K \times 3/5 - (4K+7K+2K+10K)*2}{2} \times 0.273}{1.05} = -0.52 \tag{17}$$

$$V_{D_3} = \frac{\frac{70 - (2K+7K+1K+10K)*2}{2} \times 0.8179}{1.05} = 11.68 \tag{18}$$

$$V_{D_1+D_2} = \frac{\frac{70K - (10K + 15K + 4K + 7K + 2K + 10K) * 2}{2} \times 0.233}{1.05} = -2.88 \tag{19}$$

These results clearly stablish as favorite an individual dataset that covers the requirements, instead of the combination of two dataset.

# 6 Conclusion

This paper has presented a practical, strategy-aligned framework for data valuation and monetisation that bridges the long-standing gap between technical data-quality assessments and market-oriented pricing. Leveraging the four-perspective structure of the BSC and integrating it with a comprehensive taxonomy of over one hundred metrics and KPIs, our model enables organisations to translate high-level strategic imperatives, whether cost leadership, differentiation, or focused innovation, into quantifiable value drivers. The incorporation of multi-criteria decision-making via the ANP further tailors each valuation, ensuring that governance, cost, quality, and utility dimensions receive weightings aligned with actual business priorities.

Our hybrid methodology synthesises internal valuation techniques (cost- or income-based, utility-oriented, and qualitative scoring) with external market benchmarks to produce transparent, defensible value proxies for both internal investment decisions and external transactions. The three-stage workflow, initial screening, systematic metric selection and normalisation, and final value computation incorporating demand forecasts and an information-cost fraction, offers clear guidance even for organisations at early stages of data maturity. The accompanying [11] web tool operationalises these concepts, supporting survey-based pairwise comparisons and automated priority vector calculations.

Looking ahead, empirical validation across diverse industry use cases will be essential to refine parameter settings and test robustness under varying regulatory, technological, and market conditions. Future work should also investigate dynamic pricing mechanisms that adjust in real time to shifts in demand and data network effects. By equipping stakeholders with a modular, transparent, and evolution-ready valuation architecture, we aim to accelerate the transition from ad hoc, initiatives to enterprise-wide, value-centric data strategies that unlock the full economic potential of information assets.

# 7 Acknowledgments

This work was conducted with the financial support of the EU's Horizon Digital, Industry, and Space program under grant agreement 101092989, and by Research Ireland under Grant No. XXXXXX, which is co-funded under the European Regional Development Fund. For the purpose of Open Access, the author has applied a CC BY public copyright license to any Author Accepted Manuscript version arising from this submission.